# Influence of Magnetic Force on the Flow Stability in a Rectangular duct

Rahman Anisur[1], Wenqia Xu[1], Kunhang Li[1], Hua-Shu Dou[1,*], Boo Cheong Khoo[2] and Jie Mao[3]

[1] *Faculty of Mechanical Engineering and Automation, Zhejiang Sci-Tech University, Hangzhou, 310018, China*

[2] *Department of Mechanical Engineering, National University of Singapore, Singapore 119260, Singapore*

[3] *School of Mechanical Engineering, Hangzhou Dianzi University, Hangzhou, 310018, China*

**Abstract.** The stability of the flow under the magnetic force is one of the classical problems in fluid mechanics. In this paper, the flow in a rectangular duct with different Hartmann ($Ha$) number is simulated. The finite volume method and the SIMPLE algorithm are used to solve a system of equations and the energy gradient theory is then used to study the (associated) stability of magnetohydrodynamics (MHD). The flow stability of MHD flow for different Hartmann ($Ha$) number, from $Ha=1$ to 40, at the fixed Reynolds number, Re=190 are investigated. The simulation is validated firstly against the simulation in literature. The results show that, with the increasing $Ha$ number, the centerline velocity of the rectangular duct with MHD flow decreases and the absolute value of the gradient of total mechanical energy along the streamwise direction increases. The maximum of $K$ appears near the wall in both coordinate axis of the duct. According to the energy gradient theory, this position of the maximum of $K$ would initiate flow instability (if any) than the other positions. The higher the Hartmann number is, the smaller the $K$ value becomes, which means that the fluid becomes more stable in the presence of higher magnetic force. As the Hartmann number increases, the $K$ value in the parallel layer decreases more significantly than in the Hartmann layer. The most dangerous position of instability tends to migrate towards wall of the duct as the Hartmann number increases. Thus, with the energy gradient theory, the stability or instability in the rectangular duct can be controlled by modulating the magnetic force.

*Corresponding author.
  *Email addresses:* huashudou@yahoo.com

# 1 Introduction

The instabilities in magnetohydrodynamics (MHD) flows have obtained much concern since it would affect the product quality of magnetic casting, stirring, and metallurgy [1]. The duct flow of an electrically conducting fluid with an imposed constant magnetic force is nearly optimal for analyzing fundamental properties of turbulence in liquid metal MHD as well as implications for many technological processes. Other applications include the continuous casting of steel or self-cooled liquid metal blankets for nuclear fusion reactors. Despite the simple geometrical setup, the flow presents several key effects: turbulence with mean shear and the Hartmann boundary layers at the walls, respectively, perpendicular and parallel to the magnetic force.

Hartmann and Lazarus [2] experimented on the flow of mercury in a homogeneous magnetic force with various aspect ratios of rectangular ducts, and investigated the changes in the skin friction and the suppression of turbulence caused by magnetic force. Brouillette and Lykoudis [3] carried out experiment on the MHD turbulent flow in a rectangular duct 5:1 aspect ratio with insulated walls, and investigated the laminarization effect under a uniform and strong magnetic force. The skin friction coefficient $C_f$ is observed to be a function of $Hr(\equiv Ha/\mathrm{Re}\times 10)$.

Gradner and Lykoudis [4] studied turbulent pipe flow in a transverse magnetic force and found the reduction of the skin friction coefficient was remarkably observed at around $Hr=30$ ($R=333$) ($R=\mathrm{Re}/Ha$ instead of $Hr$). Sajid et al [5] investigated the non-similar analytic solution for MHD flow and heat transfer in a third-order fluid over a stretching sheet. It is found that the skin friction coefficient decreases as the magnetic parameter increases or the third grade parameter increases. Ishak et al [6] worked on the mixed convection boundary layer in the stagnation-point flow of an incompressible viscous fluid over a stretching vertical sheet. Kobayashi [7] performed the large eddy simulation (LES) of the MHD turbulent channel flow employing the Smagorinsky model (SM), the Dynamic Smagorinsky model (DSM) and the Coherent Structure model (CSM). It shows that the CSM is able to predict higher transition Hartmann number much better than the DSM. Kobayashi [8] did LES study on turbulent MHD duct flows with a uniform magnetic field perpendicular to insulated walls. It was found that the coherent structures near the Hartmann layer are suppressed more than those near the sidewall with the increasing magnetic effect. Grandet et al [9], Takhar and Ram [10], and Duwairi and Damseh [11] studied the hydromagnetic flow in MHD. Pantokratoras [12] also investigated the MHD boundary layer flow over a heated stretching sheet with variable viscosity, and good results have been achieved.

Tinsober [13] made use of the electromagnetic force generated by the interaction electrodes and magnetic poles to make clear that the electromagnetic force is conducive to improving the stability of the boundary layer. Takhar el al. [14] studied the numerical solution to the MHD stability problem for dissipative Couette flow in a narrow gap under some condition, and find that the effect of the magnetic field is to inhibit the onset of instability, it being more so in the presence of conducting walls than in the presence of non-conducting walls. Kim and Lee [15] performed stability analysis of a viscous, incompressible and electrically conducting liquid sodium flow in an annular linear induction electromagnetic pump for sodium coolant circulation of a Sodium Fast Reactor (SFR) when transverse magnetic fields permeate the sodium fluid across the narrow annular gap. It was shown that a magnetic field has a significant stabilizing effect on the liquid sodium flow.

The understanding of the mechanism of instability and transition in MHD flows is almost exclusively limited to Hartmann layer rather than sidewall layer. Recently, Lingwood and Alboussière [16] performed linear stability analysis of Hartmann layer, and a critical Reynolds number was determined to be 48250*Ha*, which are two orders of magnitude higher than observation in experimental studies. Hossain and Khan [17] investigated the effect of moving wall on the tube and magnetic field on hydrodynamic stability. Dong et al. [18] investigated the case of a square cross-section duct with Re=5000 in a preliminary study, and find that the mean flow is stable at either large or small Hartmann number, but unstable at a moderate value *Ha=30*.

In recent years, Dou and co-authors [19-23] proposed an energy gradient theory with the aim to clarify the phenomenon of flow instability as well as the onset of transition from laminar to turbulent flow applicable to the wall-bounded shear flow.

In this study, the magnetohydrodynamics (MHD) duct flow is investigated. The energy gradient theory is used to provide the theoretical analysis of the instability of the MHD duct flow. The mechanism leading to flow instability for the rectangular duct is discussed, and the dominating parameter for these phenomena is given.

## 2 Numerical Model and Analytical Method

### 2.1 Numerical model

The laminar flow of an electrically conducting fluid in a rectangular duct in an external magnetic force is numerical simulated. The cross section of the duct is measured as *2b* (*2b=40*mm) by *2a* (*2a=50*mm), and the length is $z_0=600$mm as shown in Fig. 1. For the given Hartmann number ($\equiv LB \times (\sigma/\rho\upsilon)^{1/2}$) is taken to be, *Ha=10*.

As shown in Fig. 1, the characteristic length *L=a*, is the half of the height of the rectangular duct. A uniform constant magnetic field *B=0.007119*T is imposed. The magnetic force is parallel to the *y*-axis, as shown in Fig. 1. (The boundary layer which is parallel with the magnetic force is called parallel layer, whereas the boundary layer which is perpendicular to the magnetic force is called Hartmann layer). The walls of the duct are assumed to be rigid and insulated. The fluid medium is the liquid metal NaK in which material properties and other physical parameters are provided in Table 1.

Table 1: Physical parameters and material properties

| Physical parameters and material properties | | |
|---|---|---|
| Rectangular width | 2b | 40 mm |
| Rectangular height | 2a | 50 mm |
| Axial length of laminar model | $z_0$ | 600 mm |
| Fluid density | ρ | 868.2 Kg.m$^{-3}$ |
| Viscosity coefficient | υ | $1.05 \times 10^{-6}$ m$^2$.s$^{-1}$ |
| Fluid conductivity | σ | $2.878 \times 10^{6}$ s.m$^{-1}$ |
| Applied magnetic field | B | 0.007119T |

In the simulation, the boundary conditions for the inlet section is set as the average inlet mean velocity of $w_0=0.008$ms$^{-1}$. The outlet boundary condition is the pressure outlet, and the flow Reynolds number is Re ($\equiv w_0 L/\upsilon$) is 190.

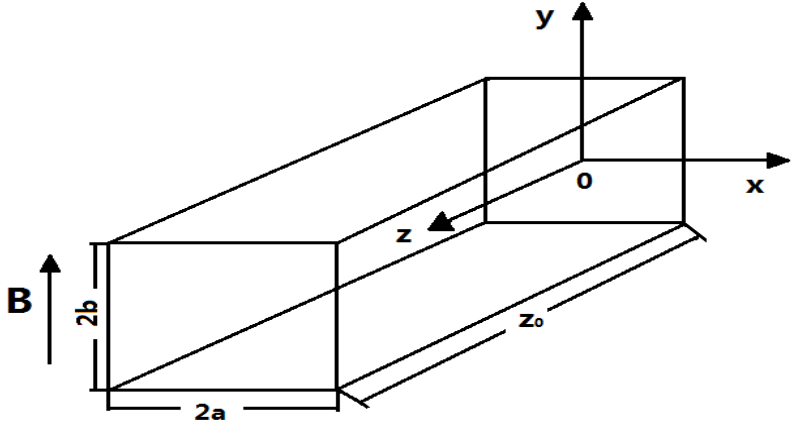

Figure 1: Computational domain of the 3D rectangular duct.

The simulation for the incompressible MHD flow is based on the finite volume method, and SIMPLE algorithm, in which the following equations namely the continuity equation, Navier-Stokes equation with the Lorentz force, the generalized Ohm's law and Maxwell equation at low Reynolds number [24] are solved. These are respectively:

$$\nabla \cdot \mathbf{u} = 0 \tag{1}$$

$$\frac{\partial \mathbf{u}}{\partial t} + \mathbf{u} \cdot \nabla \mathbf{u} = -\nabla p + \nu \nabla^2 \mathbf{u} + \frac{1}{\rho} \mathbf{j} \times \mathbf{B} \tag{2}$$

$$\mathbf{j} = \sigma(-\nabla \phi + \mathbf{u} \times \mathbf{B}) \tag{3}$$

$$\nabla \cdot \mathbf{j} = 0. \tag{4}$$

Here, $\mathbf{u}$ is the velocity, $p$ is pressure, $\mathbf{j}$ is the current density, $\mathbf{B}$ is the magnetic field, and $\phi$ is the electric potential.

## 2.2 Brief of energy gradient theory

The energy gradient theory was proposed based on the Newtonian mechanics [19-23]. Whenever a fluid particle is agitated, it will commence a wavering movement. From the classical theory of the Brownian movement, the fluid particle exchanges vitality and energy all the time in collision. The agitated fluid particle will collide with other particles in the transverse direction even as it streams along its streamline direction. In doing so, this particle would acquire energy expressed as $\Delta E$ say after several cycles. At the meantime, the particle would dissipate energy in its motion along the streamline given by $\Delta H$. There exists a critical value of the ratio between $\Delta E$ and $\Delta H$ such that, above which the particle would leave its original path and moves into another streamline (with higher energy or lower energy) and below which the particle would continue on its (original) streamline with its waving motion. Referring to [19-23], we can express the criteria of stability as follows:

$$K\frac{u'_m}{u} < Const. \tag{5}$$

$$K = \frac{\partial E/\partial n}{\partial E/\partial s} \tag{6}$$

Here, $E = p + 0.5\rho u^2$ is the total mechanical energy (total pressure) per unit volumetric fluid, $s$ is along the streamwise direction (In this study, $z$ is along the streamwise direction), $n$ is along the transverse direction, $u$ is the streamwise velocity, and $u'_m$ is the amplitude of the disturbance of the velocity.

## 3 Results and Discussions

### 3.1 Distributions of mean flow

Figure 2 shows the distribution of the total velocity at five different values of *Ha* number at the cross-section plane *z=0.5*. It can be seen that, as the Hartmann number increases the velocity at the center of the cross-section decreases.

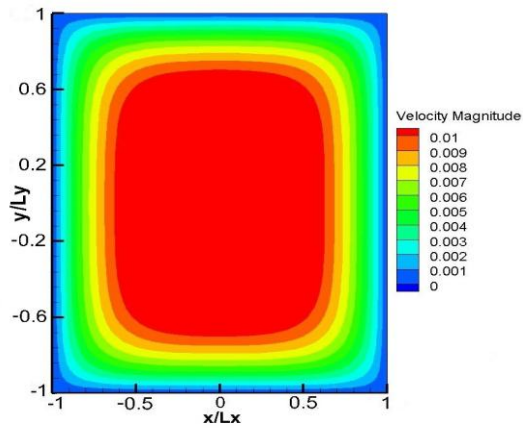

(a) *Ha=1*

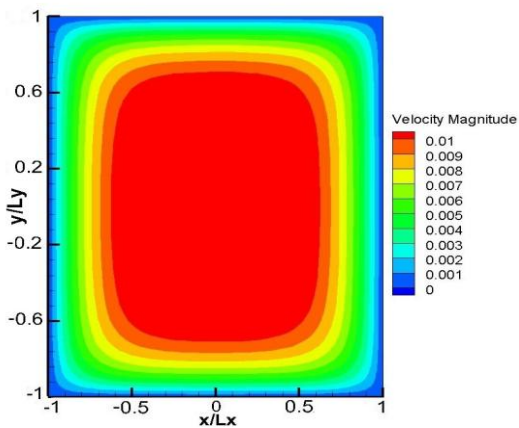

(b) *Ha=5*

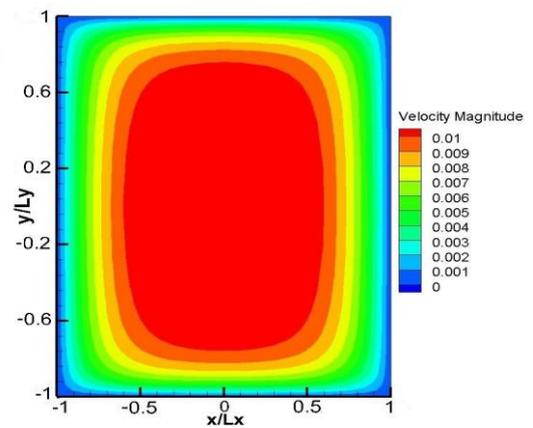

(c) *Ha=10*

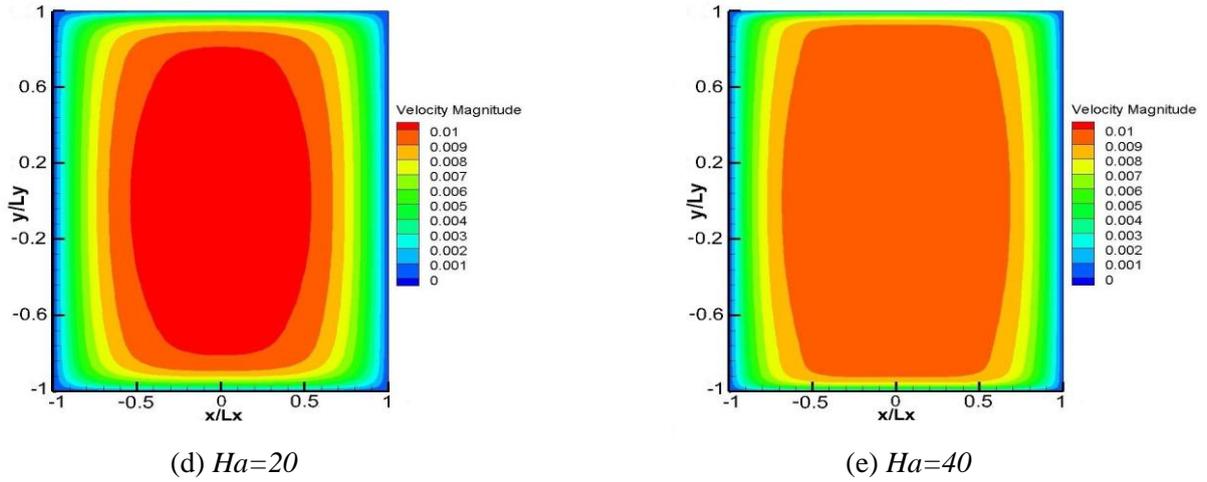

(d) *Ha=20*      (e) *Ha=40*

Figure 2: Distribution of total velocity at different *Ha* number at the cross-section plane of *z=0.5* at Re=190.

In Fig. 3, the distribution of velocity at *x=0* for *Ha=10* at the cross-section *z=0.5* is given. The result is validated with the simulation of Hao et al [25]. Good agreement is obtained.

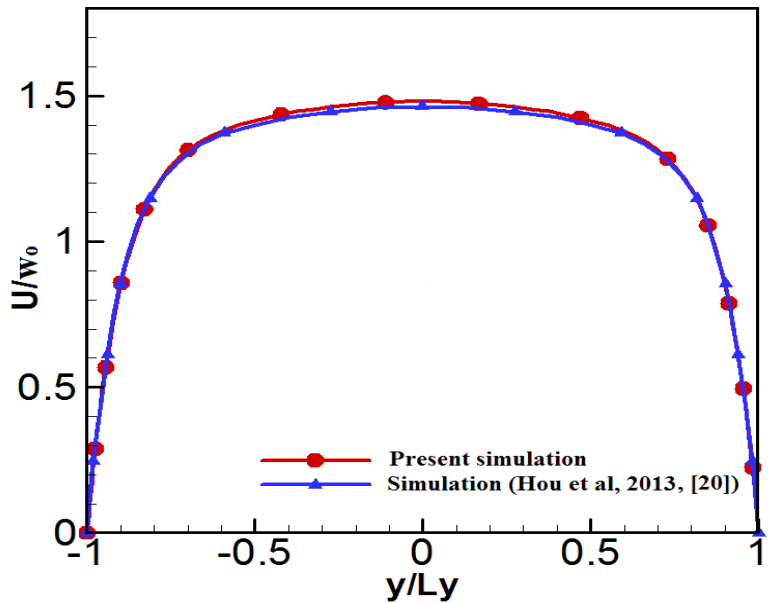

Figure 3: Validation of the computed results with Hou et al [25] simulation results at *Ha=10* and Re=190.

Figure 4 shows the electric current density at the cross-section of *z=0.5* for the different Hartmann number. It may be reiterated from Fig. 1 that the magnetic force is in the *y*-axis direction and the conductive fluid flows in the *z*-axis direction which is traversing the magnetic force. According to the Fleming left hand rule, the electric current will then flow in the negative *x*-axis direction towards the duct wall; the flow changes its direction accordingly. The induced current in the conductive fluid circulation is symmetrical on both sides of the horizontal center line. With the increase of Hartmann number, the density of electric current also tends to increase. The increasing electrical current density hence pushes the center of the two symmetrical circular paths towards the wall. The velocity profile is influenced by the increasing current density, which can be seen from Fig. 2 and Fig. 5.

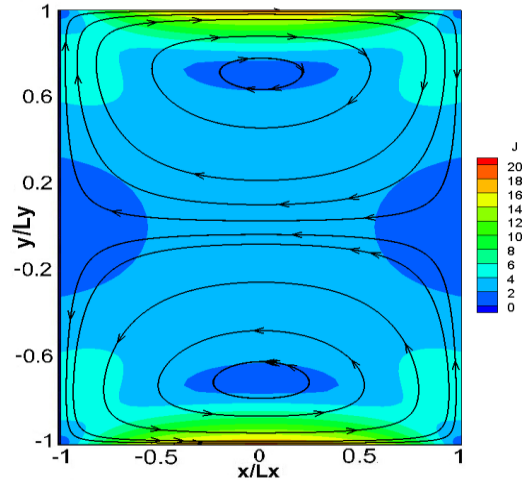

(a) *Ha=1*

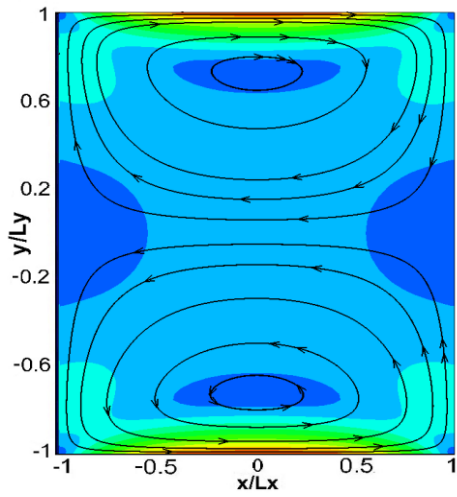

(b) *Ha=5*

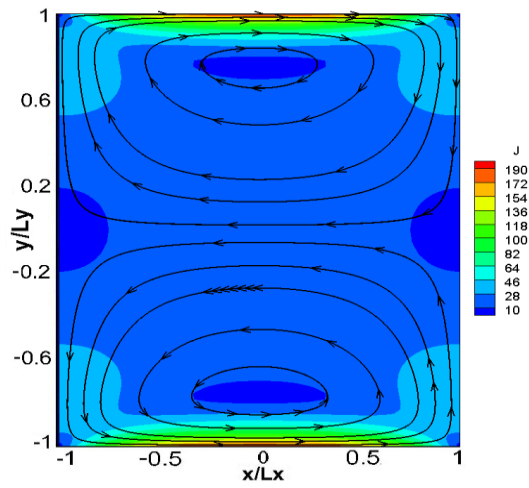

(c) *Ha=10*

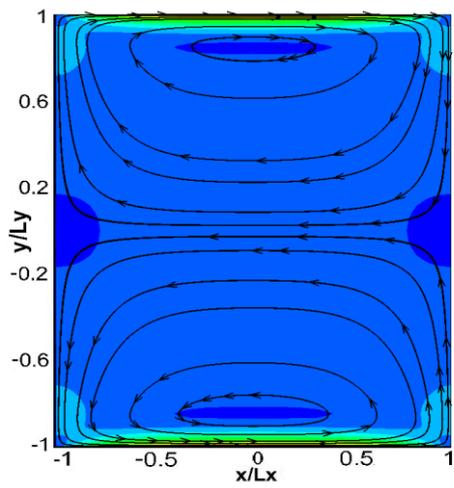

(d) *Ha=20*

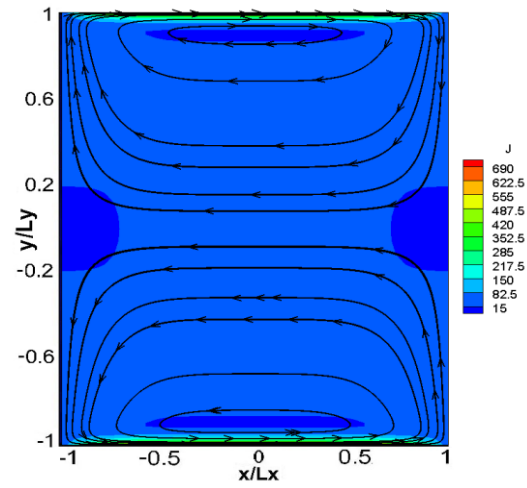

(e) *Ha=40*

Figure 4: Distributions of electric current density *J* at different *Ha* number at the cross-section plane of *z=0.5*.

Figure 5 shows the comparison of velocity distribution at *y=0* and *x=0*, respectively, at the cross-section plane of *z=0.5* for the different Hartmann number. It can be seen that when the Hartmann number increases from *Ha=1* to *Ha=40*, the centerline velocity decreases, see Fig. 5(a). According to the Fleming left hand rule, since the direction of the current is in the negative *x*-axis direction as shown in Fig. 4, the direction of the Lorentz force will then be in the negative *z*-axis direction. It is noted that the direction of the Lorentz force is opposite to the direction of fluid motion, so it hinders the movement of the main flow. Along the *x*-direction near the wall, both the symmetric current paths are almost parallel with the magnetic force, so the value of Lorentz force near the wall is almost zero. As such, the velocities near the wall do not change much. When the Hartmann number increases from *Ha=1* to *Ha=40*, the velocity of the center region along the *y*-axis direction decreases about the same as the *x*-axis direction, which attributed to the Fleming left hand rule. In the region near the wall along the *y*-axis, the current direction is in the *x*-axis direction, the Lorentz force will then be the same direction as the fluid motion, which leads to the increase of velocity near the wall of the *y*-axis; see Fig. 5(b), which shows a good agreement with Brouillette and Lykoudis [3].

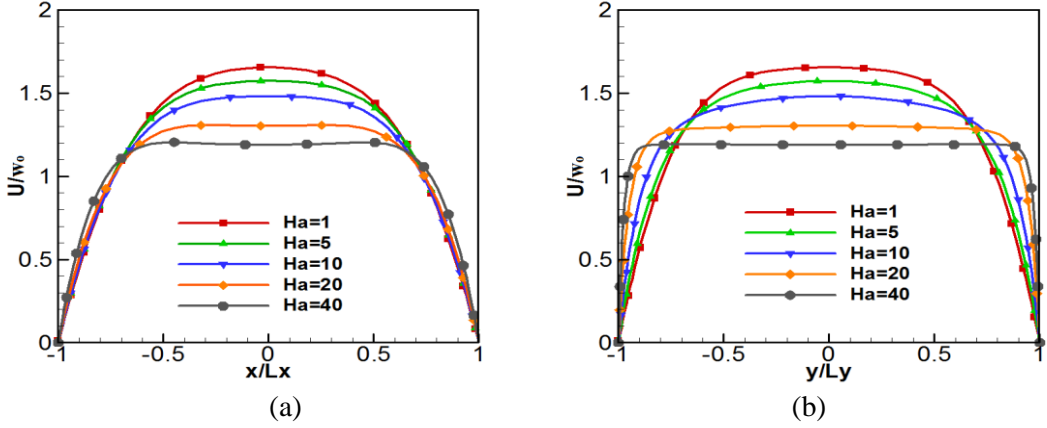

(a)            (b)

Figure 5: Distribution of axial velocity at the cross-section plane of *z=0.5* under different *Ha* number. (a) The axial velocity along *x* direction at *y=0*; (b) The axial velocity along y direction at *x=0*.

### 3.2 Discussion on flow stability under magnetic force

Figure 6 shows the distribution of the energy gradient function *K* at the cross-section plane of *z=0.5* at Re=190. The imposed magnetic force is in the *y*-direction. With the increase of Hartmann number from (a) *Ha=1* to (e) *Ha=40*, the value of *K* decreases along both axis. And the maximum of the energy gradient function $K_{max}$ moves towards the wall in the *y*-axis direction. According to the energy gradient theory, as the *K* value decreases, the MHD flow will become more stable. Therefore, as the Hartmann number increases, the flow will become more stable because the *K* value is decreasing. This is consistent with the fact that the magnetic force can enhance the stability of the flow.

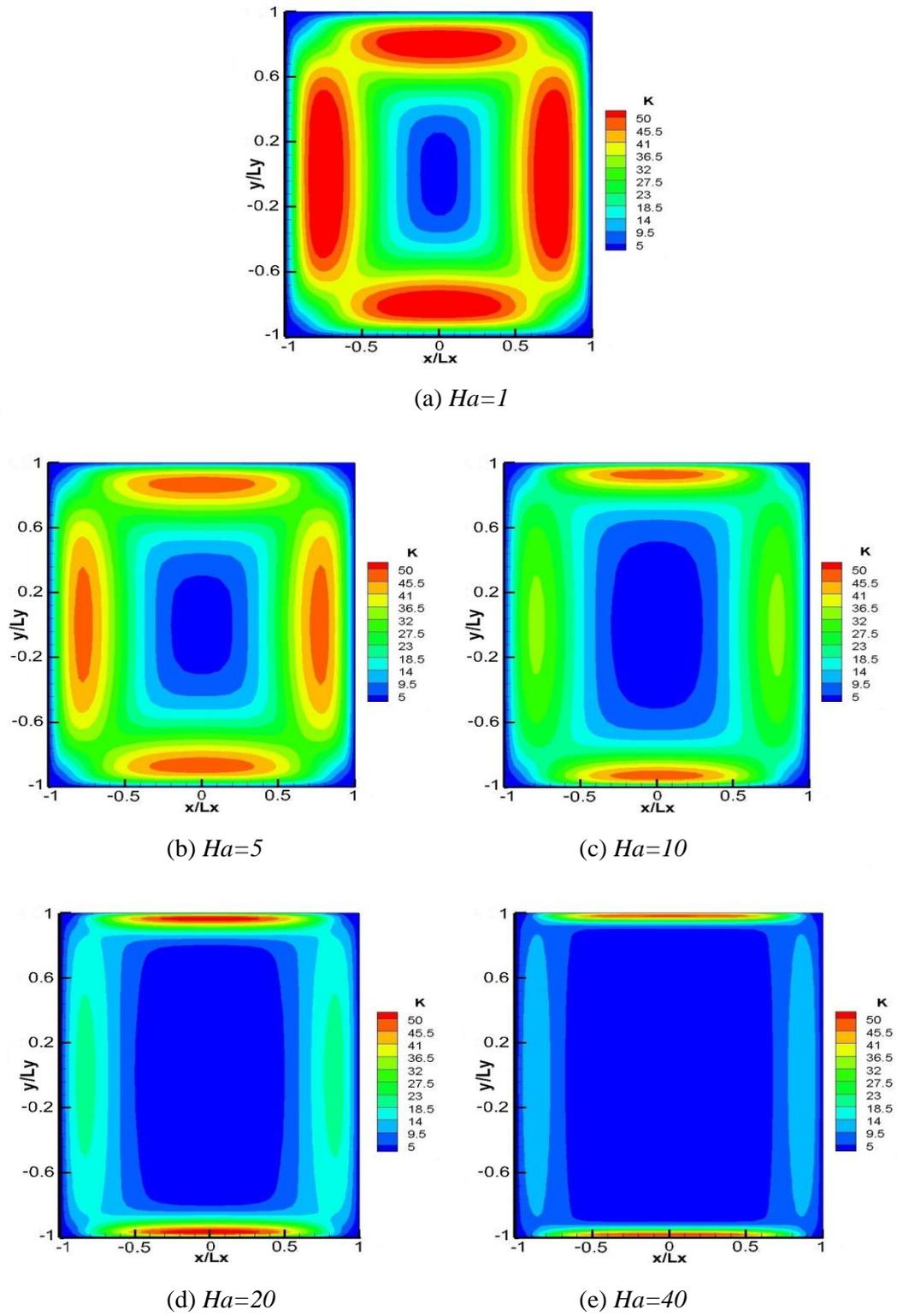

(a) *Ha=1*

(b) *Ha=5*     (c) *Ha=10*

(d) *Ha=20*    (e) *Ha=40*

Figure 6: Distribution of *K* at different *Ha* number at the cross-section plane of *z=0.5*.

Figure 7 shows the distribution and the comparison of dimensionless function $K$ for all the MHD flow along both $x=0$ and $y=0$ at the cross-section plane of $z=0.5$. It can be seen that the dimensionless function $K$ near the wall attains its maximum at Hartmann number $Ha=1$ along both $x$-axis and $y$-axis. In Fig. 7, there are two values of $K_{max}$ along both the $x$-axis and $y$-axis. The position of the $K_{max}$ is the place where the oscillation and disturbance first occur. When the Hartmann number is increased from $Ha=1$ to $Ha=40$, the value of $K_{max}$ along the $x$-axis decreases gradually (i.e. decreased oscillation and disturbance) and moves toward the wall gradually, and the flow becomes more stable than before. Again along the $y$-axis, when the Hartmann number is increased from $Ha=1$ to till $Ha=40$, it shows that the value of $K_{max}$ decreases only slightly and also moves towards the wall gradually. The results of the dimensionless function $K$ or $K_{max}$ shows that, the higher the magnitude of magnet force, the more stable the flow is.

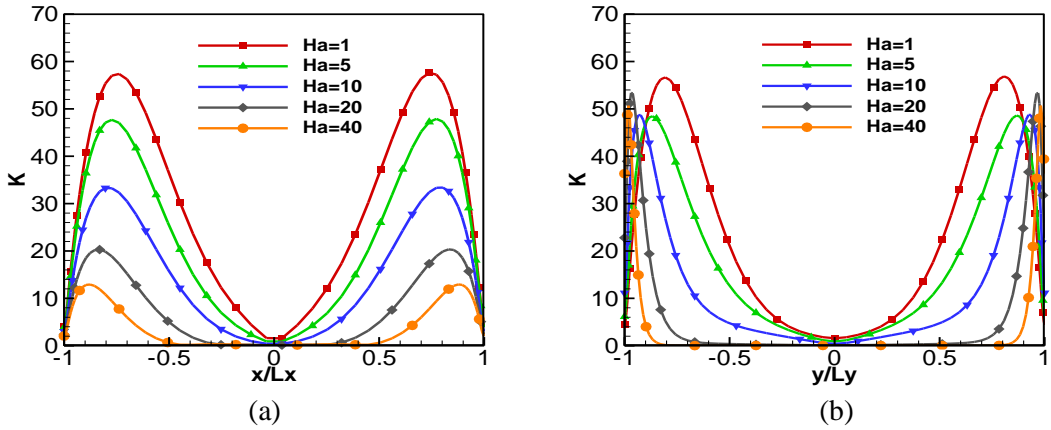

Figure 7: Distribution of $K$ at the cross-section plane of $z=0.5$ under different $Ha$ number. (a) $K$ along $x$ direction at $y=0$; (b) $K$ along y direction at $x=0$.

Figure 8 shows the distribution of $dE/dn$ at $y=0$ and $x=0$ at the cross-section plane of $z=0.5$. At $y=0$, as the Hartmann number increases, the position with the maximum value of $dE/dn$ moves towards the wall along the $x$-axis. As mention above, the Lorentz force near the wall along the $x$-axis is almost zero. As such, the maximum value of $dE/dn$ under the different Hartmann number near the wall is almost the same. The velocity of the centerline along the $x$-axis decrease gradually. Because the Lorentz force, which opposes the direction of the fluid motion of the centerline along the $x$-axis, so the $dE/dn$ near the centerline also decreases. Figure 9 shows the distribution of $dE/dz$ at $y=0$ and $x=0$ at the cross-section plane of $z=0.5$. From the distribution along $x$-axis at $y=0$, the absolute value of $dE/dz$ increases as the Hartmann number increases. From the analysis of $dE/dn$ and $dE/dz$, we can further see that the $K$ value is decreasing while the maximum value of $K$ moves towards the wall gradually along the $x$-axis as the Hartmann number increases.

From Figure 8, with the increasing Hartmann number, the value of current gradient along the normal direction of the velocity at the centerline at the $y$-axis tends to be smaller, whereas the current gradient along the normal direction of the velocity near the wall becomes larger. With the increased Hartmann number, the position with the maximum value of $dE/dn$ increases and tends to move towards the wall gradually.

From the distribution of *dE/dz* along the *y*-axis in Fig. 9(b), the absolute value of *dE/dz* also increases as the Hartmann number increases. From the analysis of *dE/dn* and *dE/dz*, the *K* value along the *y*-axis nearly keeps to a stable quantity and the maximum value of *K* moves towards the wall gradually as the Hartmann number increases.

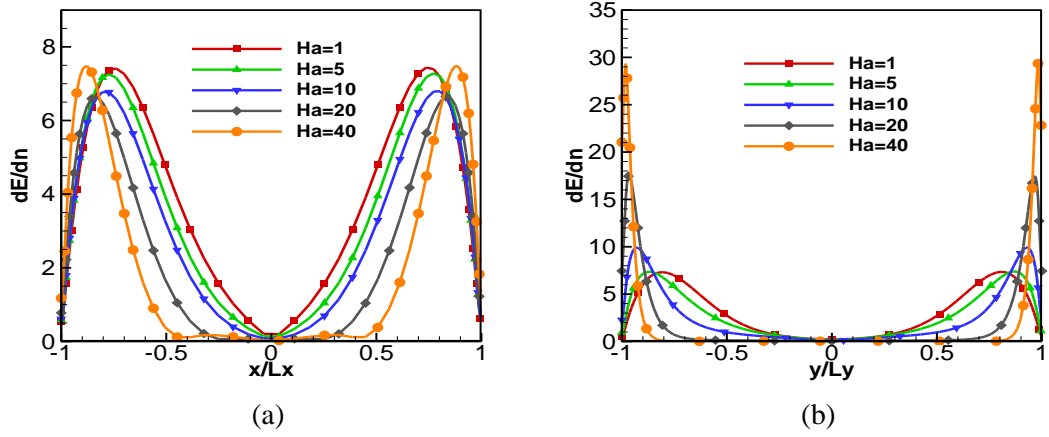

(a)                    (b)

Figure 8: Distribution of *dE/dn* at the cross-section plane of *z=0.5* under different *Ha* number. (a) The *dE/dn* along *x* direction at *y=0*; (b) The *dE/dn* along *y* direction at *x=0*.

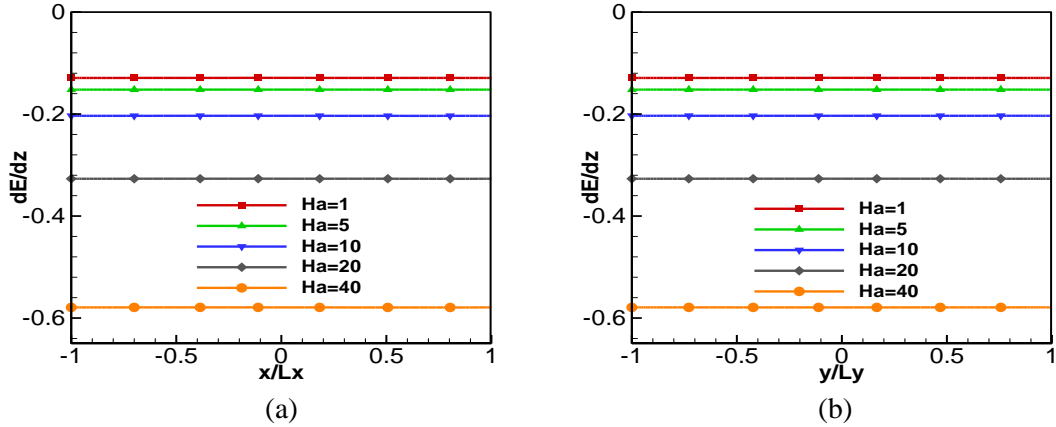

(a)                    (b)

Figure 9: Distribution of *dE/dz* at the cross-section plane of *z=0.5* under different *Ha* number. (a) The *dE/dz* along x direction at *y=0*; (b) The *dE/dz* along y direction at *x=0*.

In summary, at y=0, at the cross-section plane of z=0.5, as the Hartman number increases, the *K* value near the wall along the x-axis decreases and the maximum value of *K* moves towards the wall gradually.

At *x=0*, at the cross-section plane of *z=0.5*, the *K* value along the *y*-axis keeps to nearly a constant quantity and the maximum value of *K* moves towards the wall gradually as the Hartmann number increases. Above all, as the Hartman number increases the *K* value decreases. This means that by adding the magnetic force, we may improve the stability of the flow.

# 4 Conclusions

This paper presents numerical investigation and theoretical study of magnetohydrodynamics (MHD) flow in a rectangular duct to analyze the flow instability via the energy gradient theory. The energy gradient function $K$ is employed to characterize the stability of fluid flow. The conclusions are summarized as follows:

As the Hartmann number increases, the centerline velocity in the rectangular duct tends to decrease. The velocity gradient in the Hartmann layer increases significantly, but it does not vary in the parallel layer significantly.

As the Hartmann number increases, the absolute value of the gradient of total mechanical energy along the streamwise direction, $dE/dz$, increases. Thus, the magnetic fluid enhances the drop of total pressure in streamwise direction.

The higher the Hartmann number is, the smaller the $K$ value becomes, which means that the fluid flow becomes more stable with higher magnetic force. The $K$ value in the parallel layer decreases more significantly than the Hartmann layer as the Hartmann number increases. Thus the parallel layer is more stable than the Hartmann layer.

The most dangerous position of instability according to the energy gradient theory tends to migrate towards the wall of the duct as the Hartmann number increases.

The results of analysis with the energy gradient theory obtained agreement with the simulations and experiments in literature. Therefore, the stability (or instability) in the rectangular duct may be controlled by modulating the magnetic force in the flow field with the energy gradient theory.

**Acknowledgments**

This work is supported by National Natural Science Foundation of China (51536008, 51579224), Zhejiang Province Science and Technology Plan Project (2017C34007), and Zhejiang Province Key Research and Development Plan Project (2018C03046).